\documentclass[twocolumn,showpacs,preprintnumbers,amsmath,amssymb]{revtex4}
\usepackage{graphicx}
\usepackage{dcolumn}
\usepackage{bm}

\def\bd{\begin{document}} \def\ed{\end{document}}
\def\bmp{\begin{minipage}} \def\emp{\end{minipage}}
\def\bcc{\begin{center}} \def\ecc{\end{center}}     \def\npg{\newpage}
\def\beq{\begin{equation}} \def\eeq{\end{equation}} \def\hph{\hphantom}
\def\be{\begin{equation}} \def\ee{\end{equation}} \def\r#1{$^{[#1]}$}
\def\n{\noindent} \def\ni{\noindent} \def\pa{\parindent}
\def\hs{\hskip} \def\vs{\vskip} \def\hf{\hfill} \def\ej{\vfill\eject}
\def\cl{\centerline} \def\ob{\obeylines}  \def\ls{\leftskip}
\def\underbar#1{$\setbox0=\hbox{#1} \dp0=1.5pt \mathsurround=0pt
   \underline{\box0}$}   \def\ub{\underbar}    \def\ul{\underline}
\def\f{\left} \def\g{\right} \def\e{{\rm e}} \def\o{\over} \def\d{{\rm d}}
\def\vf{\varphi} \def\pl{\partial} \def\cov{{\rm cov}} \def\ch{{\rm ch}}
\def\la{\langle} \def\ra{\rangle} \def\EE{e$^+$e$^-$} \def\pt{p_{\rm t}}
\def\Yct{Y_{\rm cut}} \def\Rct{R_{\rm cut}} \def\dct{d_{\rm cut}}  
\def\kt{k_{\rm t}} \def\HH{hadron-hadron} \def\yct{y_{\rm cut}}  
\def\bitz{\begin{itemize}} \def\eitz{\end{itemize}}
\def\btbl{\begin{tabular}} \def\etbl{\end{tabular}}
\def\btbb{\begin{tabbing}} \def\etbb{\end{tabbing}}
\def\beqar{\begin{eqnarray}} \def\eeqar{\end{eqnarray}}
\def\\{\hfill\break} \def\dit{\item{-}} \def\i{\item}
\def\bbb{\begin{thebibliography}{9} \itemsep=-1mm}
\def\ebb{\end{thebibliography}} \def\bb{\bibitem}
\def\bpic{\begin{picture}(260,240)} \def\epic{\end{picture}}
\def\akgt{\noindent{Acknowledgements}}
\def\fgn{\noindent{\bf\large\bf Figure captions}}
\newcommand{\JETSET}{{\sc jetset}{\small 7.4}} 
\newcommand{\HERWIG}{{\sc herwig}{\small 5.9}}
\newcommand{\SPPS}{{\sc s}{\small p$\bar{\rm p}$}{\sc s}}
\newcommand{\JADE}{{\sc jade}} \newcommand{\DURHAM}{{\sc durham}}
\newcommand{\LUND}{{\sc lund}} \newcommand{\ISR}{{\sc isr}} 
\newcommand{\QCD}{{\sc qcd}} \newcommand{\NFM}{{\sc nfm}}
\newcommand{\LEP}{{\sc lep}} \newcommand{\CERN}{{\sc cern}}
\newcommand{\ALEPH}{{\sc aleph}} \newcommand{\OPAL}{{\sc opal}}
\newcommand{\DELPHI}{{\sc delphi}} \newcommand{\Lthree}{{\sc l}{\small 3}}
\newcommand{\UAFIVE}{{\sc ua}{\small 5}}
\newcommand{\MC}{{\sc m}{\small onte} {\sc c}{\small arlo}}
\newcommand{\Fermilab}{{\sc f}{\small ermilab}}
\newcommand{\QuantumChromoDynamics}{{\sc q}{\small uantum}{\sc c}{\small hromo}
{\sc d}{\small ynamics}}   \def\z{\hskip3pt}
\newcommand{\TEVATRON}{{\sc tevatron}} 
\newcommand{\Brookhaven}{{\sc b}{\small rookhaven}}
\newcommand{\RHIC}{{\sc rhic}} \newcommand{\LHC}{{\sc lhc}}
\bd

\title{On the Phase Space Partition in High Energy Collisions}

\thanks{This work is supported in part by NSFC under project
19975021.}

\author{Liu Lianshou}

\affiliation{Institute of Particle Physics, Huazhong 
Normal University, Wuhan 430079 China}

\email{liuls@iopp.ccnu.edu.cn}

\begin{abstract}
In high energy hadron-hadron and \EE\z collisions, to isolate a part of the 
phase space in multi-hadron final states
is necessary for exploring the underlying dynamics. It is shown 
that the partition of phase space according to the value of rapidity, 
popularly used in hadron-hadron collisions, is inappropriate for the 
study of \EE\z collisions. The proper way in the latter case is to
identify the visible jets and take them as objects for detailed study,
forming an extended phase space. The value $\dct^{(0)}$ of the 
distance-measure for the identification of visible jets is determined.
A new variable $r$ is introduced to 
further partition the phase space inside jets, which possesses a very
good anomalous scaling property, showing that the $r$-distribution inside
jets is a self-similar fractal.  
\end{abstract}


\maketitle
The dynamics of multi-hadron final states in
high energy collisions is an idea laboratory for the study of the perturbative
and nonperturbative properties of the basic theory of strong interaction
--- \QuantumChromoDynamics\z \QCD. This study, starting from the sixties 
of last century, has a long history. The multiplicity as well as the 
phase space volume have increased for orders of magnitude since then.
It becomes necessary to take a part of the phase space for detailed analysis 
instead of to merely study it as a whole. 

In \HH\z collisions, already in the 1970\hskip0.5pt s the \ISR\z experiments 
started to study the ``central rapidity region'' $|y| < \Yct$~\cite{ISR}, 
where the rapidity $y$ is defined as 
\beqar  
y=\frac{1}{2}\ln\frac{E+p_\parallel}{E-p_\parallel},
\eeqar
with $E$ the energy and $p_\parallel$ the longitudinal momentum of the
particle in the c.m. frame, $\Yct$ is a specially chosen cut
smaller than the maximum rapidity reached in the collisions. 
The longitudinal direction is taken to be the 
direction of the momenta of the two incident particles.

Later, the successful operation of the p-$\bar{\rm p}$ collider \SPPS\z
at \CERN\ raised the c.m.  energy $\sqrt s$ for an order of magnitude.   
It enabled the \UAFIVE~\cite{UA5} and other collaborations 
to carry on their study in various rapidity windows.

More recently, this approach has been extended to \EE\z collisions. 
For example, \DELPHI~\cite{DELPHI} and \OPAL~\cite{OPAL} Collaborations 
have studied the intermittency and correlations in 
hadronic Z$^0$ decay in a restricted rapidity region $|y|<2$, 
with the thrust or sphericity axis taken as the longitudinal direction.

However, the dynamics of \EE\z collisions is different from 
that of \HH\z collisions. 
In \HH\z collisions the central rapidity region has a special physical
meaning. Particle production occurs mainly in this region. The regions
outside of the central one are the fragmentation regions, where the
leading particle effect is present. On the other hand, in \EE\z collisions
the final-state system comes from a point source (virtual photon or Z$^0$).
There is no leading particle effect, or, if some similar effect exists in
the individual jets, it is not directed
along the thrust (or sphericity) axis in multi-jet
events, see e.g. Fig.1 and Fig.2. 
In this case, to study the ``central rapidity region'' with thrust or 
(sphericity) axis as longitudinal direction is doubtful.

In 1994 J. D. Bjorken~\cite{Bj} proposed to 
isolate the gluon jets for the study of the
crucial features of the underlying dynamics
and put forward the idea of ``extended phase-space'',
cf. figure~1 of Ref.\cite{Bj}. In this approach gluon jets produced in \EE\z
collisions are first identified and then taken out as a part of the extended
phase-space. Further partition of phase space is done inside the jets.

Different ways of cutting phase-space, as e.g. those done in 
Ref.~\cite{DELPHI}\cite{OPAL} 
and those proposed in Ref.~\cite{Bj}, will apparently
result in different physics. What is the proper way for the
partition of phase-space?
How to realize it? are urgent problems not only for the study of \EE\z
collisions but also for the study of the ultra-high energy \HH\z and/or
relativistic heavy ion collisions at \Fermilab-\TEVATRON, \Brookhaven-\RHIC\z 
and \CERN-\LHC, where a vast amount of jets are being or will be produced.   

In this letter we will try to give an  answer to these questions.
For illustration, we will use \LUND\ \MC\ \JETSET\ to generate a sample of
1.5 million \EE collision events at $\sqrt s=91.2$ GeV. 

\begin{figure}
\includegraphics[width=9.5cm]{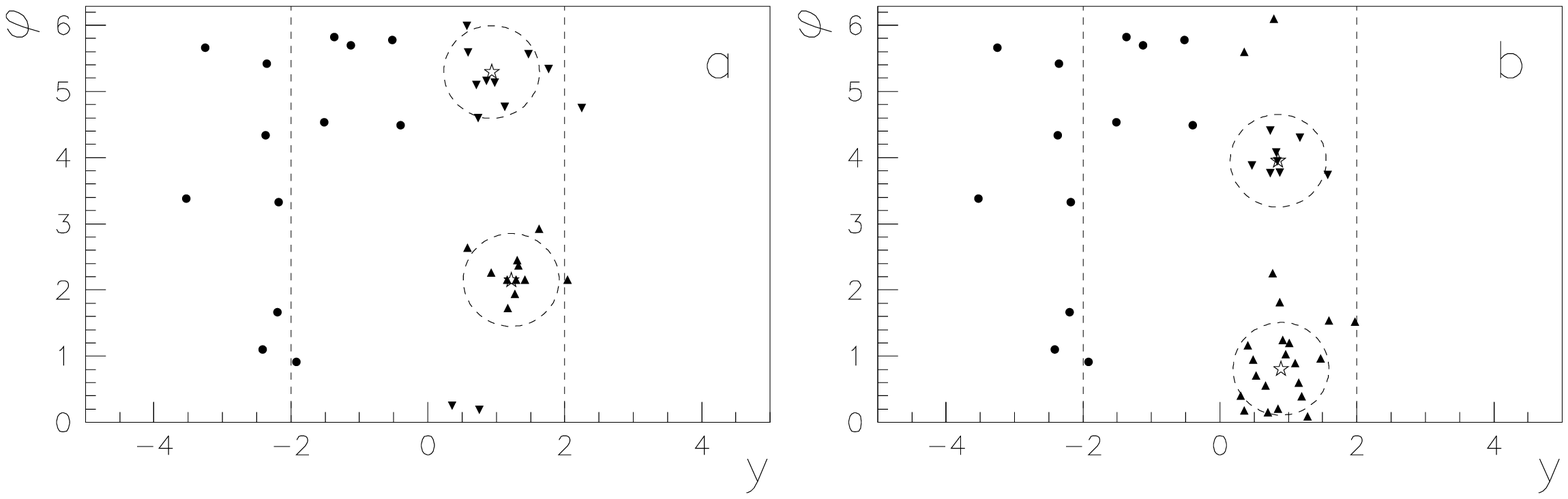}
\caption{\label{fig:lego} The lego plot of two typical 3-jet events 
determined by \DURHAM\z algorithm with log\z $\dct =-2.2$.
Downward and upward triangles are the particles of the two right-hand-side 
jets, respectively. Open stars are the corresponding jet axes.
Dashed circles have radius $r=0.7$.}
\end{figure}

Let us first analyse the usefulness of ``central rapidity region'' in \EE\z
collisions. As already pointed out, the final state hadrons
in these collisions are produced from a point source. There is no leading
particle effect, or in multi-jet events such effect is not along the 
``longitudinal'' (thrust or sphericity) direction. 
In these collisions there are 2-jet, 3-jet, 4-jet, ...
events. A cut in rapidity (with the thrust or sphericity axis as
"longitudinal") will have very different effect on different-number-of-jet 
events. For example, in a 2-jet event a $y$ cut will cut out the most
energetic particles in both jets; while in a 3-jet event a $y$ cut will only 
cut out the most energetic particles of one jet but retain almost all the
particles of the other two jets. Two typical examples are given in Fig.1,
where the dotted lines denote the rapidity cut $|y|<2$. It cuts out a large
part of the left-hand-side jet but retains almost all the particles of the 
other two jets. The results from such an asymmetric cut is hard to be
interpreted.

Physically, a more appropriate approach is not to select a group of
particles with rapidity
in some range but to isolate the individual jets. The corresponding 
cut condition is then $r<\Rct$~\cite{Bj} instead of $y<\Yct$, 
where~\cite{ftnt01}
\beqar
 r=\sqrt{(\eta-\eta_{\rm jet})^2 + (\vf-\vf_{\rm jet})^2},
\eeqar
$\eta$ and $\vf$ are the pseudo-rapidity and azimuthal angle of a final 
state particle in a jet, respectively, while $\eta_{\rm jet}$ and 
$\vf_{\rm jet}$ are those of the jet.  $\Rct$ is a number around unity. 
For the dashed circles in Fig.1 this number has been chosen to be
$\Rct=0.7$~\cite{Bj}. Note that in Eq.(2) we have used pseudo-rapidity
$\eta$ defined as
\beqar
 \eta= -\ln \tan ({\theta}/{2})
\eeqar
instead of the rapidity defined in Eq.(1). At high energy,
$\eta \approx y$. For the study of phase space
partition, $\eta$-$\vf$ is a natural choise, which are directly connected
with the corresponding variables in a space-time description~\cite{Bj}.
Note also that, when a jet has its axis (the total momentum of the particles 
inside it) nearly collinear with the thrust (or sphericity) axis, which 
is used as longitudinal to define the variables $\eta$ and $\vf$, the 
azimuthal angle $\vf$ of the particles inside the jet 
will be spreading and the jet will not be visible in the $\eta$-$\vf$ 
lego-plot, cf. the left-hand-side jets of Fig.1 and Fig's.2 ($a$), ($c$).
Therefore, in the following we will restrict ourself to consider only those 
jets whose axes have an angle greater than 15$^\circ$ and less than 165$^\circ$
with the thrust (or sphericity) axis. Note that the gluon jets, coming from
hard gluon radiation, will most probably satisfy this condition.

In this approach the basic questions are:
How to identify the jets? How to partition the phase space further
after isolating out the individual jets?
Let us try to answer these questions.

The identification of jets from the final-state multiparticle system can be
done using some algorithm, e.g. the \JADE~\cite{Jade} or 
the \DURHAM~\cite{Durham} ones. In these schemes a distance measure is defined
for counting the separation between two jets (or two particles). 
For example, in the \DURHAM\ algorithm this measure is defined as 
\beqar
 d_{ij}=\frac{2\min(E_i^2,E_j^2)}{s}(1-\cos \theta_{ij}),
\eeqar
which is related to the relative transverse momentum $\kt$ as~\cite{kt} 
\beqar
 (\kt)_{ij} = \sqrt{d_{ij}} \cdot \sqrt{s}.
\eeqar
In Eq.(4)  $E_i$ and $E_j$ are the energies of the two jets 
(or two particles), $\theta_{ij}$ is the angle between them and
$s$ is the square of c.m. energy of the event. 
Note that we have used the notation $d_{ij}$ instead of $y_{ij}$, currently 
used in the literature, to avoid confusion
with the rapidity cut. Jets or particles with $d_{ij}\leq \dct$
are combined into a single jet. This procedure is
repeated until all pairs of jets $i$ and $j$ satisfy $d_{ij}>\dct$.

In principle, $\dct$ could be chosen arbitrarily. The resulting ``jets'' are,
however, very different for different values of $\dct$.  
For our present purpose, we want to cut the phase-space according to jets,
so the ``jets'' should be directly observable, cf. Fig's.1 and 2. 
Such directly observable jets will be referred to as ``visible jets''. 

It has been shown in Ref.~\cite{LCHPrd} that the scale for visible jets is
$5\leq \kt \leq 10$ GeV. For the case of $\sqrt s=91.2$ GeV, which we are 
taking as example, this range of $\kt$ 
corresponds to $10^{-2.5}\leq \dct\leq 10^{-1.9}$.

\begin{figure}
\includegraphics[width=6cm]{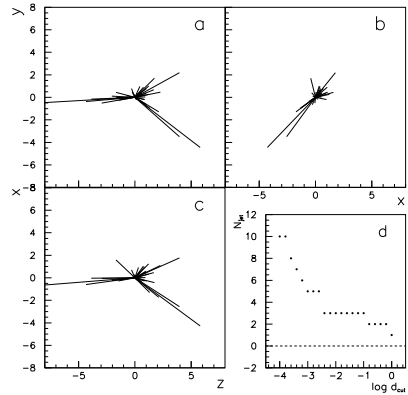}
\caption{\label{fig:pxyz} ($a$)-($c$) The same event as in Fig.1($b$) plotted
in momentum space, in $z$-$y$ ($a$), $x$-$y$ ($b$) and $z$-$x$ ($c$) 
planes, respectively. ($d$) The number of jets in this event
determined by different values of $\dct$.}
\end{figure}

In order to be more apparent, we plot in Fig's.2 ($a$)--($c$) 
the view in momentum 
space of the same event, shown in Fig.1($b$) as lego-plot.
The number of jets $N_{\rm jet}$ versus
log\z $\dct$ in this event is given in Fig.2($d$). It can be seen from the
figure that if we were using a value of $\dct$ as small as $10^{-4}$ (or
$10^{-3}$) then we would conclude that this is a 10-jet (or 5-jet) event; 
while if we used a value of
$\dct$ greater than $10^{-1}$ then we would take this event as a 2-jet one.
Although all these values of $\dct$ are equally possible in principle, 
it is very unlikely that 
there are 10 (or 5) jets or only 2 jets in this event, cf. Fig's.1 and 2. 
For our present purpose of constructing extended phase space through 
plotting sub-lego plots of individual jets, this event should certainly be  
considered as a 3-jet event.
It can be seen from Fig.2($d$) that a value
of $\dct$ about equal to or a little smaller than $10^{-2}$ will combine the 
wrongly identified five jets into three. This is the $\dct$ value 
corresponding to visible jets.

\begin{figure}
\includegraphics[width=4cm]{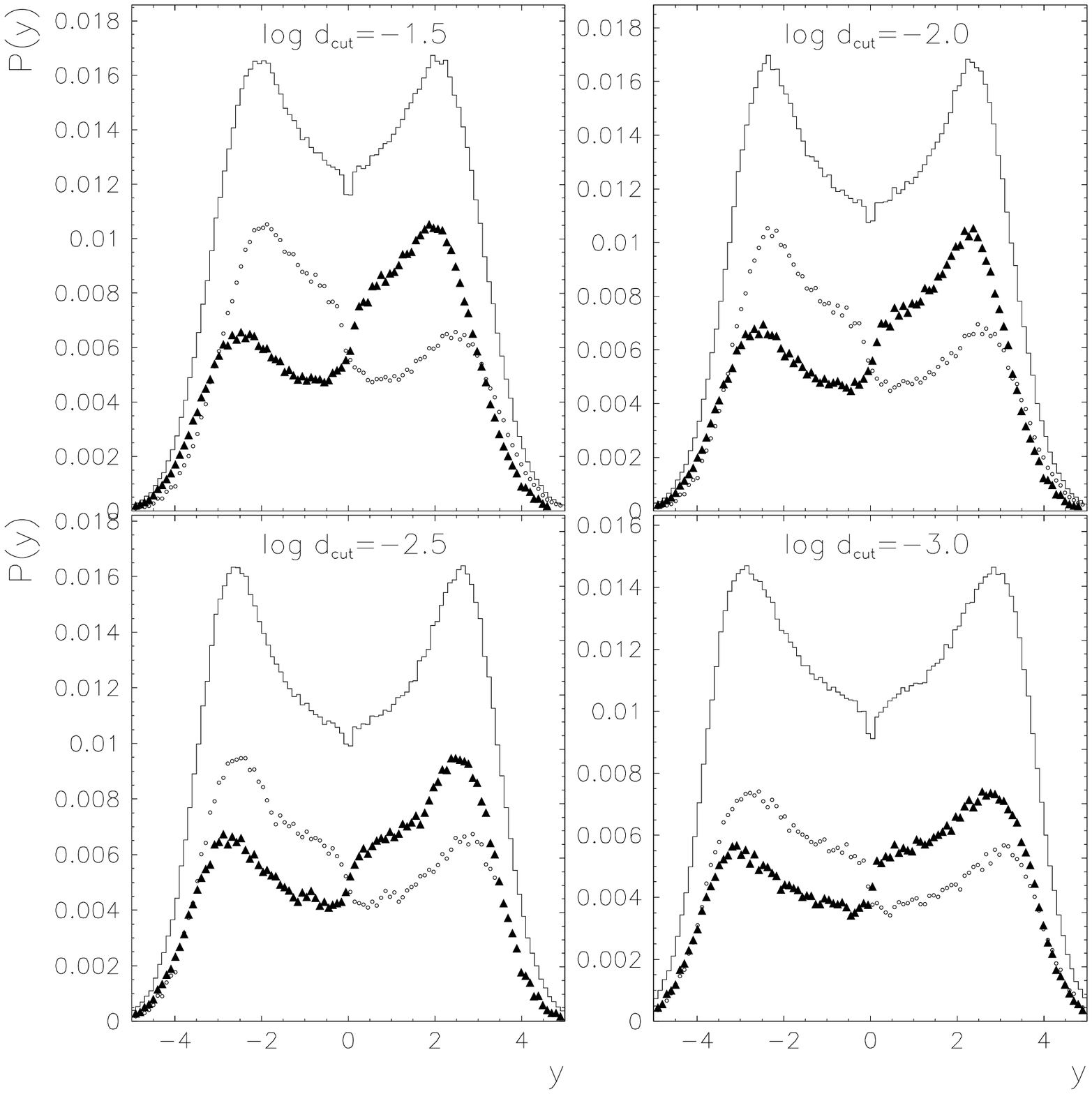}
\caption{\label{fig:disy} The rapidity distributions (histograms) of 2-jet 
events determined by \DURHAM\ algorithm with different values of $\dct$.
Full triangles --- that of the events with the bigger jets at right.
Open circles --- that of the events with the bigger jets at left.}
\end{figure}

A further check of this statement could be obtained from the rapidity
distributions of 2-jet events for four different values of $\dct$, shown
in Fig.3 as histograms.
In the same figures, the full triangles (open circles) are the distributions 
for those events in which the bigger jet is at the right (left). It can be 
seen from the figures that for log\z$\dct=-2.5\sim -2.0$ the full triangles 
manifest a shoulder at about $y=1.8$, showing that at this value of 
$\dct$ a third jet starts to appear inside the big jet.
This confirms the assertion that log\z$\dct= -2.5 \sim -1.9$ is the appropriate
distance-measure for identifying the visible jets in \EE\ collisions at
$\sqrt s=91.2$~GeV. For convenience, we will
in the following refer to this measure as $\dct^{(0)}$ and use a value
$\dct^{(0)}=10^{-2.2}$. 

Now let us turn to the second question: How to partition the phase space 
further inside the identified jets?

The first step is to take out the identified jets and draw sub-lego plots for
them, forming an extended phase space together with the mother lego plot. 
In these sub-lego plots the jet axes are taken as the center to define 
new variables 
\beqar
  \eta' =  \eta-\eta_{\rm jet} , \quad \vf' = \vf-\vf_{\rm jet} .
\eeqar
As example, the sub-lego plots of the two identified jets in Fig.1($b$)
are shown in Fig.4.

\begin{figure}
\includegraphics[width=6cm]{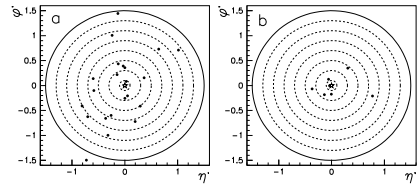}
\caption{\label{fig:sublego} The sub-lego plots of the two identified jets 
in Fig1($b$).}
\end{figure}

\begin{figure}
\includegraphics[width=7cm]{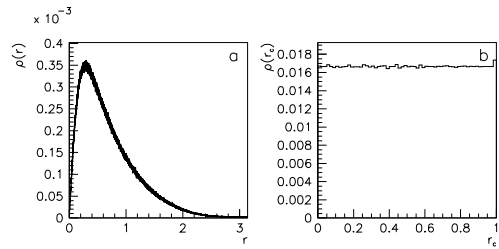}
\caption{\label{fig:rhor} ($a$) The distribution of $r$ and ($b$) that of the
corresponding cumulant variable $r_{\rm c}$.} 
\end{figure}

The pair of variables
($\eta_{\rm jet}$, $\vf_{\rm jet}$) determine the direction of jet
axis. A circular symmetry is expected to exist
around this point in the $\eta$-$\vf$
plane. Therefore, a natural variable for the description of the phase space 
region inside a jet is the distance $r$ in $\eta$-$\vf$ plane from the
particle inside the jet to the jet axis ($\eta_{\rm jet}$, $\vf_{\rm jet}$), 
cf. Eq.(2). 

The distribution of $r$ is shown in Fig.5($a$). It has a pick at about 
$r=0.3$ and tends to zero when $r$ increases. Choosing a certain value of
$R_0$, e.g. $R_0=1.5$, shown as solid circle in Fig's.4, a phase space
region can be defined as $0<r\leq R_0$. A partition of this region using
different values of $r$ results in a series of rings, cf. the dashed circles
in Fig's.4. The distance $\delta r$ between the neighboring rings is the 
scale of the partition. 

It is interesting to check whether there is a scale independence or
anomalous scaling in the 
1-dimensional $r$ space, or in other words, whether the $r$-distribution is 
a self-similar fractal. This can be done using the standard 
technique~\cite{kittel}.

\begin{figure}
\includegraphics[width=6cm]{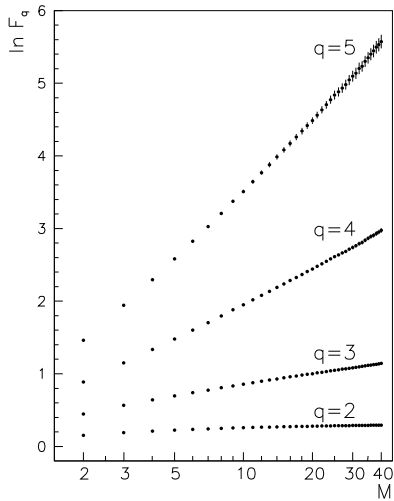}
\caption{\label{fig:scaling} The log\z$F_q$ versus log\z$M$ 
plot.}
\end{figure}

The first step is to change the variable $r$ into the corresponding
cumulant variable $r_{\rm c}$ defined as~\cite{cumulant}
\beqar
 r_{\rm c} = {\int_0^r \rho(r)\d r}\f/ {\int_0^{R_0} \rho(r)\d r}\g. 
\eeqar
to eliminate the influence of nonflat distribution. The resulting distribution
of $r_{\rm c}$ is shown in Fig.5($b$).

The second step is to divide the region $0<r\leq R_0$ 
($0<r_{\rm c}\leq 1$) into $M$ bins and
define the $q$th order normalized factorial moments \NFM\ as~\cite{BP}
\beqar
  F_q(M) =
    \frac{1}{M}\sum\limits_{m=1}^{M}
    \frac{\la n_m(n_m-1) \cdots (n_m-q+1)\ra}
    {\la n_m \ra ^q  },
\eeqar
where $n_m$ is the number of particles in the $m$th bin, $\la\cdots\ra$
denotes the average over the sample of identified jets. The advantage of
\NFM\ is that it can eliminate the statistical fluctuations provided the
latter is of the Poinsonian type~\cite{BP}. The anomalous scaling of \NFM\
\beqar
 F_q(M) \propto M^{\phi_q} \quad M\to \infty
\eeqar 
as the increasing of the partition number $M$ or as the 
decreasing of the phase space scale $\delta r = R_0/M$, 
will signal the existence of fractal property --- self-similarity.

In Fig.6 is shown the log\z$F_q$ versus log\z$M$ plot in the region 
$0<r\leq \pi$~\cite{ftnt02}
for $q=2,3,4,5$. 
It can be seen that very good scaling is obtained for $M=2$ -- 40.
This astonishing phenomenon is closely related to the color coherence
--- angular ordering in parton fragmentation and is worthwhile further
investigation, both experimentally and theoretically.

In this letter the proper way for the phase space partition in high energy
\EE\z collisions is analysed in some detail.   It is shown 
that the partition of phase space according to the value of rapidity, 
popularly used in hadron-hadron collisions, is inappropriate for the 
study of \EE\z collisions. The proper way in the latter case is to
identify the visible jets and take them as objects for detailed study,
forming an extended phase space. The value of the distance-parameter 
$\dct$ of the \DURHAM\z algorithm, which is able to identify the 
visible jets, is discussed in detail and is found to be about 
$\dct^{(0)}=10^{-2.2}$ (at c.m. energy $\sqrt s=91.2$ GeV).

A new variable $r$ is introduced to 
further partition the phase space inside jets, which possesses a very
good anomalous scaling property, showing that the $r$-distribution inside
jets is a self-similar fractal. This spectacular phenomenon, being connected
with the underlying dynamics, is worthwhile further theoretical and 
experimental investigations.

\vskip1.5cm

\begin{acknowledgments}
The author thanks Wu Yuanfang, Wolfram Kittel, Yu Meiling and Chen Gang 
for helpful discussions.
\end{acknowledgments}

\def\J#1#2#3#4{{#1} {\bf #2}, {#3} (#4)}
\def\PRL{\it Phys. Rev. Lett.} \def\PRep{\it Phys. Rep.}
\def\PR{\it Phys. Rev.} \def\ZP{\it Z. Phys.} \def\JP{\it J. Phys.}
\def\NP{\it Nucl. Phys.}  \def\PL{\it Phys. Lett.}
\def\EPJ{\it Eur. Phys. J.}

\ed
\begin{thebibliography}{9}
\bibitem{ftnt01} 
The range of $\vf$ is [0,$2\pi$]. If $|\vf-\vf_{\rm jet}|>\pi$
then $r=\sqrt{(\eta-\eta_{\rm jet})^2 + (2\pi-|\vf-\vf_{\rm jet}|)^2}$.
\end{thebibliography}
